\documentclass[pop,twocolumn,showpacs]{revtex4-1}
\usepackage[colorlinks=true,citecolor=blue]{hyperref}
\usepackage{dcolumn}
\usepackage{bm}
\usepackage{epsfig}
\usepackage{hyperref}
\usepackage{empheq}
\usepackage{amssymb,amsmath,commath}
\usepackage{romannum}

\begin{document}
\pagenumbering{arabic}

\title{
A Nonlinear Dynamics Characterization of The  \\
Scrape-off Layer Plasma Fluctuations} 

\author{A. Mekkaoui}%
  \affiliation{Institute for Energy and Climate Research - Plasma Physics,
 Research Center J\"ulich GmbH, Association FZJ-Euratom, D-52425 J\"ulich, Germany}
   \email{s.mekkaoui@fz-juelich.de}

\begin{abstract}
A stochastic differential equation for the plasma density dynamics is derived,  consistent with the
experimentally measured distribution and the theoretical quadratic nonlinearity.
The plasma density is driven
by a multiplicative Wiener process and evolves on the turbulence correlation time scale, while the linear growth is 
quadratically damped by the fluctuation level. The sensitivity of intermittency to the nonlinear dynamics 
is investigated by analyzing the Langevin representation of two intermittent distributions, showing the agreement 
between the quadratic nonlinearity and the gamma distribution.

\end{abstract}
\pacs{02.50.Ey, 52.65.Ff, 05.10.Gg, 52.25.Gj}
\maketitle
The scrape-off layer (SOL) plasma of magnetic confinement fusion devices exhibits a universal intermittent behavior characterized
by a strongly non-Gaussian statistics
and spatiotemporal correlations \cite{Carreras_2005}. The intermittent character of fluctuations is closely 
related to a bursty convective transport of over-density coherent structures (blobs) and edge localized
 mode filaments (ELMs) \cite{Ippolito_2011}.   
Basically the deviation from normality is caused by the quadratically nonlinear turbulence 
, coupling the plasma density
to the electric potential,
i.e. $\partial_t \psi +\dots \propto \psi\psi$  \cite{Krommes_2002}. For instance an assumed Gaussian 
initial condition $\psi_0$ evolve to a Chi-square random variable (r.v.) ($\psi_0^2$) and so on.\\
\indent The closure procedure reducing
the coupled deterministic turbulence equations to a nonlinear Langevin 
equation is still a challenging problem \cite{Krommes-08}, although a substantial 
theoretical effort has been devoted to the statistical characterization of intermittency in a turbulent media.  
In climatology science \cite{Lenschow}  as well as in fusion plasma research \cite{Sandberg-2009},
an attempt to the dynamics interpretation of intermittency and its statistical signature have been addressed by 
describing the intermittent quantity $\psi$ as a quadratic polynomial of a Gaussian variable $g$, i.e. 
\begin{equation}\label{Lensc}
\psi(t)=g(t)+\omega g^2(t),
\end{equation}
where $\omega$ is the non-normality parameter which measures the deviation from the Gaussian statistics
 and the strength of the nonlinear coupling. 
It is not surprising that the process Eq.~(\ref{Lensc}) satisfies the universal parabolic 
relation between the Kurtosis and the Skewness, i.e. $K=aS^2+b$, 
observed in several turbulent media \cite{Krommes-08}, because 
the variable $\psi+1/4\omega$ is distributed as $\omega\left(g+1/2\omega \right)^2$, which is the non-central 
Chi-square distribution by construction, and is a particular 
case of the gamma distribution measured in the edge of fusion devices  and satisfying $K=1.5S^2+3$ \cite{Graves05,Labit}. 
Unfortunately the underlying  physical mechanism of Eq.~(\ref{Lensc}) is still 
not clear and the Gaussianity assumption of the dynamical variable $g$ is too strong.\\ 
\indent Based on a physical intuitions, several other
 existing stochastic models are able to explain the emergence of the gamma statistics from a turbulent media. 
In their investigation of scattered radiation from a fluctuating background, Jakeman {\it et al.} 
\cite{Jakeman_1977} showed that the gamma probability distribution function (PDF) appears 
as a limit distribution of a finite sum of independent and identically distributed random perturbations $x_i$, 
 \begin{equation}\label{Jak}
  X(t)=\sum_{i=1}^{N} x_i,
 \end{equation}
 where their number $N$ obeys a birth-death-immigration process with a respective rates $b$, $d$ and $m$ 
\cite{Jakeman_1977}. Without any condition 
on the perturbers distribution $P(x_i)$, and when the death rate is close to the birth rate ($b \simeq d$),
 then $X$ is gamma distributed 
with $\mu=\left<x\right>$ scale factor and $\nu=m/b$ shape factor,
\begin{equation}\label{Eqgamma}
  P(X)=\dfrac{X^{\nu-1}}{\mu^{\nu}\Gamma(\nu)}\exp\left(-X/\mu\right)
 \end{equation}
On the other hand, 
in Ref. \cite{Garcia_2012}
the gamma distribution is derived by applying the Campbell's 
theorem \cite{Rice} to the plasma density signals, assumed to be a linear superposition of $K$ bursts, i.e.,
\begin{equation}\label{Garc}
 \psi(t)=\sum_{i=1}^{K} x_i F(t-t_i)
\end{equation}
with exponentially distributed intensity $P(x)=1/\mu\exp(-x/\mu)$, 
waiting time between bursts arrivals $P(t)=\exp(-t/\tau_w)$
and burst life time $F(t)=\exp(-t/\tau_d)$.  \\
\indent It is noteworthy that both stochastic processes given by Eq.~(\ref{Jak}) and Eq.~(\ref{Garc})
 lead exactly to the same PDF Eq.~(\ref{Eqgamma}), when 
the waiting time and the duration time are given by $\tau_w =m^{-1}$ and $\tau_d=b^{-1}\simeq d^{-1}$. \\
The equivalence between both derivations of the gamma statistics has a deep physical
meaning, since we could make correspondence between the immigration process and the waiting time,
and between the birth process and the duration time. 
\indent The similarities between these two gamma processes is 
extended beyond the univariate distribution by investigating
their temporal correlation.
The covariance function of the process Eq.~(\ref{Garc}) is calculated using
 a random noise properties \cite{Rice}, 
and assuming an exponential burst life time distribution,
\begin{equation}\label{cov}
C(t)=\tau_w/\tau_d\mu^2\exp(-t/\tau_d),
\end{equation}
showing that the correlations are introduced only through a single burst duration, since a given burst is 
independent of each others as in the Kubo-Anderson process \cite{Brissaud}. 
The correlation structure of the shot noise process Eq.~(\ref{Garc}) is consistent
with that of Eq.~(\ref{Jak}), which is also exponential with the inverse death rate
as  a correlation time $C(t)\propto \exp(-bt)$ \cite{Jakeman_1979}.\\
\indent The two models provide a simple physical picture of edge plasma intermittency,
where blobs are born close to the last closed flux surface (LCFS),
immigrate across the SOL convected by the cross-field velocity, 
before disappearing through dissipation and parallel transport.
Although the model Eq.~(\ref{Garc}) is consistent with the experimental measurements, it obeys a linear 
stochastic differential equation \cite{Gardiner},
what is surprising because the intermittent 
statistics in turbulent plasma is often associated 
to a nonlinear turbulence models like
the Hasegawa-Wakatani system equations \cite{Krommes_2002}. Indeed the stochastic models 
Eq.~(\ref{Jak}) and Eq.~(\ref{Garc}) could
explain the observed gamma statistics
from the response function of the instrumental devices 
point of view e.g., Langmuir probe, but their dynamical content 
is still far from the theoretical predictions.\\
\indent 
In this Letter, our purpose is to provide an insight of
the bridge between the nonlinear dynamics and the intermittent statistics in the edge plasma of magnetic fusion devices. 
Here the nonlinear dynamics is investigated starting from the experimentally measured distribution. Namely, we assume 
the gamma statistics together with a quadratic nonlinearity as a theoretical constraint
to derive a two parameters plasma turbulence equation for a given 
correlation time and fluctuation amplitude 
\begin{equation}
 R=\left(\left<\left(\psi-\left<\psi\right>\right)^2\right>\right)^{1/2}/\left<\psi\right>=\sqrt{1/\nu},
\end{equation}
where $<...>$ references to the time average. 
\noindent Indeed the gamma distribution Eq.~(\ref{Eqgamma}) obeys the Pearson equation, given in its general
 form by $\partial_\psi P(\psi)=F(\psi)P(\psi)/H(\psi)$.
The associated stochastic process is specified by the following Fokker-Planck equation for $P(\psi,t)$ \cite{Tohru},
\begin{equation}\label{eqTohru}
 \dfrac{ \partial P }{\partial t} = -
\dfrac{ \partial  }{\partial \psi}\left[ \left(  F + \dfrac{\partial H }{\partial \psi} \right) P\right]
+ \dfrac{1}{2}\dfrac{ \partial^2  }{\partial \psi^2}\left(2 H P\right).
\end{equation}
One get the following Pearson representation of the gamma distribution Eq.~(\ref{Eqgamma}),
\begin{equation}\label{eqPears}
\dfrac{\partial P(\psi)}{\partial \psi} =\dfrac{ ( 1-1/\nu) \left<\psi\right>-\psi }{  \left<\psi\right> \psi/\nu }P(\psi),
\end{equation}
where $\left<\psi\right>=\nu\mu$ being the average value of the plasma density field. Using Eq.~(\ref{eqTohru}), 
we derive the corresponding Fokker-Planck equation
\begin{equation}
\dfrac{\partial P}{\partial t}=-\dfrac{ \partial  }{\partial \psi}\left[ \dfrac{1}{\tau}\left(  \left<\psi\right>-\psi \right) P\right] +
 \dfrac{1}{2}\dfrac{ \partial^2  }{\partial \psi^2}\left[2\left<\psi\right>\psi P/\nu\tau\right] , 
\end{equation}
where we have introduced a characteristic time scale $\tau$ which is specified farther.
Then $\psi$ follows the Cox-Ingersoll-Ross process \cite{Cox_1985},
\begin{equation}\label{eqCox}
d\psi(t) = \left(  \left<\psi\right> -  \psi(t)\right)\dfrac{dt}{\tau}+\sqrt{2\left<\psi\right>\psi(t)/\nu\tau}\;dw(t),
\end{equation}
where $w$ is the Wiener process. The stochastic model Eq.~(\ref{eqCox})
  follows a root square nonlinearity because of the term $\propto\sqrt{\psi}$ and is broadly used in
 financial forecasting \cite{Lamberton}. 
The physical mechanism leading to a such dynamics is not trivial, since 
the typical turbulence models are   quadratically nonlinear \cite{Krommes_2002}. Hence the irreducible
 representation of the gamma distribution Eq.~(\ref{eqPears})  gives rise to 
some nonlinearity, but it is still not sufficient to capture the quadratically nonlinear 
dynamics expected by the plasma turbulence equations.
\indent However, Eq.~(\ref{eqPears}) is degenerate and permits one to derive a higher order nonlinear
 stochastic differential equation with the gamma PDF as a marginal.  
 \begin{figure}
 \includegraphics[scale=.45]{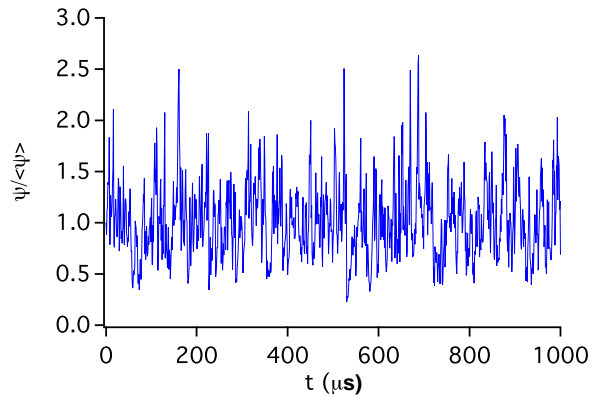}
\caption{a quiet time series of the fluctuating plasma density normalized to its local mean value, 
with $R=30\%$ and $\tau_c=3$ $\mu$s, as for a typical plasma conditions near the LCFS.}\label{Fig:1}
 \end{figure}
By multiplying the denominator 
and the numerator in the right hand side of Eq.~(\ref{eqPears}) 
by $\psi$, 
we obtain the new Pearson representation of the gamma process,
\begin{equation}
\dfrac{\partial P(\psi)}{\partial \psi} =\dfrac{ ( 1-1/\nu) \psi-\psi^2/\left<\psi\right> }{ \psi^2/\nu }P(\psi),
\end{equation}
then the corresponding stochastic differential equation follows from Eq.~(\ref{eqTohru}),
\begin{equation}\label{eqNonl}
\dfrac{\partial\psi}{\partial t} =\gamma \psi(t)-\dfrac{\psi^2(t)}{\left<\psi\right>\tau}+R\sqrt{2/\tau}\psi(t)w(t),
\end{equation}
with $\gamma=(1+R^2)/\tau$. In order to clarify the role of the time scale $\tau$, the dynamics of the correlation function $\rho(t)$ is required. 
It is straightforwardly derived from Eq.~(\ref{eqNonl}),
 \begin{equation}\label{Eq.dycor}
 \dfrac{\partial \rho(t)}{\partial t} - \gamma \rho(t) = \dfrac{\gamma}{R^2} - 
\dfrac{1}{\tau R^2 \left<\psi\right>^3} \left<\psi^2(t)\psi(0)\right>,
 \end{equation}
the term between brackets is a two-time correlation and can be approximated by,
\begin{equation}\label{Fox_app}
  \left<\psi^2(t)\psi(0)\right> = \eta\rho(t)+\kappa,
\end{equation}
 where the constant coefficients are fixed using the two particular cases $\tau_c=0$, $\rho(t)=0$ and $\tau_c=\infty$, 
$\rho(t)=1$.
 We get $\kappa=\left<\psi\right>\left< \psi^2 \right>$ and $\eta=\left<\psi^3\right>-\left<\psi\right>\left<\psi^2\right>$.
 Then Eq.~(\ref{Eq.dycor}) reduces to,
\begin{equation}
 \dfrac{\partial \rho(t)}{\partial t} + \gamma \rho(t) = 0,
 \end{equation}
the correlation function is exponential $\rho(t)=\exp(-\gamma t)$, with $\tau_c=1/\gamma$ as a correlation time.
Then $\tau=(1+R^2)\tau_c$ is nothing but the correlation time increased by a factor $1+R^2$. It is worth noticing
that the approximation Eq.~(\ref{Fox_app}) becomes exact in our case where $\psi$ is gamma distributed. This point
is not developed here and can be demonstrated 
by writing $\psi$ as a square of Gaussian r.v. and using the Wick factorization theorem.\\
\indent Equation (\ref{eqNonl}) is the main result of this Letter. It is quadratically nonlinear and evolves on a
  turbulence correlation time scale. The process Eq.~(\ref{eqNonl}) 
has a consistent physical meaning, since the linear growth is driven by the fluctuations amplitude. 
As a response to this growth, a nonlinear damping takes place and is proportional to 
the fluctuation level $\psi(t)/\left<\psi\right>$. The dynamics is driven by
 a multiplicative white noise with variance $R^2$. This stochastic drift 
disappears and the dynamics evolves deterministically when the fluctuations amplitude vanishes ($R=0$).
\indent Unlike Eq.~(\ref{Lensc}) which is based on the Gaussian assumption of the dynamical variable,
and the process Eq.~(\ref{Garc}) obeying a linear stochastic differential equation, 
our stochastic model Eq.~(\ref{eqNonl}), shows how the gamma
 distribution and the associated universal $K$-$S$ scaling rise 
self-consistently from the quadratically nonlinear dynamics of the intermittent variable. Equation~(\ref{eqNonl})  behaves like the 
stochastic generalization of the L-H transient paradigm equation \cite{Bian-10}, with the difference that the latter is quadratic
for the energy $E=((\psi-\left<\psi\right>)/\left<\psi \right>)^2$, leading to the Nakagami distribution for the plasma density.
\begin{figure}
 \includegraphics[scale=.45]{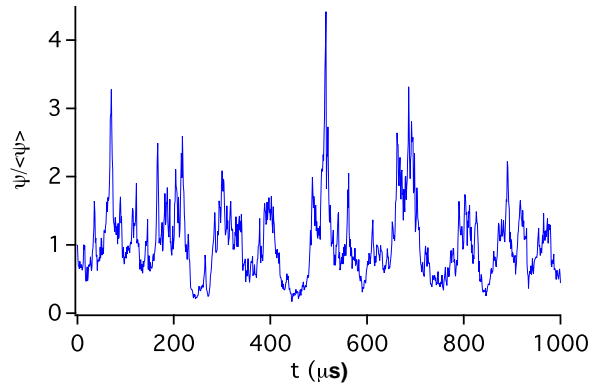}
\caption{a weakly intermittent time series of the fluctuating plasma density normalized to its local mean value, 
with $R=50\%$ and $\tau_c=16$ $\mu$s, as for a typical plasma conditions of the near SOL.}\label{Fig:2}
 \end{figure}
  \begin{figure}
 \includegraphics[scale=.45]{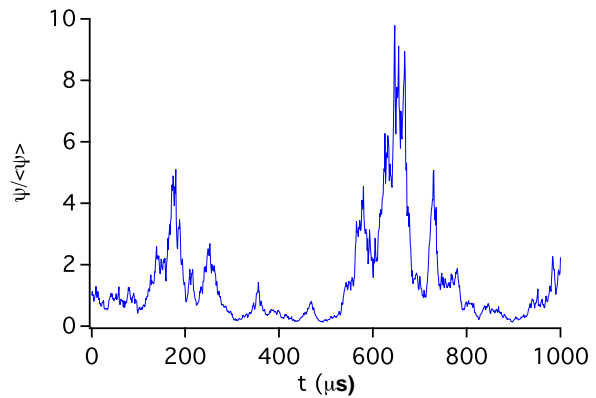}
\caption{a strongly intermittent time series of the fluctuating plasma density normalized to its local mean value, 
with $R=90\%$ and $\tau_c=55$ $\mu$s, as for a typical far SOL plasma conditions.}\label{Fig:3}
 \end{figure}
The stationary solution of Eq.~(\ref{eqNonl}) is
the gamma distribution with stationary but not necessarily homogeneous parameters.
Hence it can be used to analyze the nonlinear time series of the plasma density 
at different radial SOL positions $x$, using the local
average field $\left<\psi\right>(x)$, correlation time $\tau_c(x)$ as well as $R(x)$.
In order to simulate a plasma density time series in different conditions, 
we have used the Ito's representation of Eq.~(\ref{eqNonl}) 
for the fluctuation level $\xi=\psi(t)/\left<\psi\right>$,
 \begin{equation}\label{eqfll}
d\xi =\dfrac{\xi}{\tau_c} dt-\dfrac{\xi^2}{\tau_c(1+R^2)}dt+R\sqrt{\dfrac{2}{\tau_c(1+R^{2})}}\xi dw(t).
\end{equation}
In Fig.~\ref{Fig:1} is plotted a quiet time series of $\xi(t)$
 with time step $dt=1$ $\mu$s, the fluctuations amplitude is of $R=30\%$ and the correlation time of $\tau_c=3$ $\mu$s,
 as in the typical edge plasma conditions close to the LCFS \cite{Graves05}. Fig.~\ref{Fig:2} 
gives another time series representative of an intermediate situation corresponding to the near
 SOL with moderate fluctuations amplitude $R=50\%$ and $\tau_c=16$ $\mu$s, tree bursts exceeding
 the average value by a factor of $3$ rise in the range of $1$ ms. A strongly intermittent
 time series is plotted in Fig.~\ref{Fig:3} with $R=90\%$ and $\tau_c=55$ $\mu$s, 
tree bursts exceeding $4\left<\psi\right>$ and a super burst with the amplitude of $10\left<\psi\right>$  
are observed in the time range of $1$ ms, as is typically the case in the far SOL  \cite{Sandberg-2009}. \\
\indent In order to investigate the role of the nonlinear dynamics in the statistical characterization of
 intermittent fluctuations, we 
compare the gamma process to the existing stochastic models. The beta distribution has been used to fit the
 plasma density data in the edge of TORPEX,
 according to its capability
to have negative skewness and its finite support \cite{Labit}.   
Let consider here the standardized beta distribution,   
\begin{equation}
 P(\psi)\propto \psi^{\alpha-1}\left(1-\psi\right)^{\beta-1}, \; 0<\psi<1,\;\alpha,\;\beta>0,
\end{equation}
where $\alpha=1/R^2-\left<\psi\right>(1+1/R^2)$ and $\beta=\alpha(1/\left<\psi\right>-1)$.
 The beta distribution obeys the Pearson equation,
then using Eq.~(\ref{eqTohru}), we derive the following stochastic differential equation,
\begin{equation}
 \dfrac{\partial \psi}{\partial t}  = \dfrac{\alpha+1}{\tau}\psi+\dfrac{\beta-\alpha-3}{\tau}\psi^2  + \psi \sqrt{2\dfrac{1-\psi}{\tau}}w(t),
\end{equation}
showing that the quadratically nonlinear dynamics of the beta process is systematically accompanied with a root square nonlinearity. 
Thus unlike the beta process, the gamma process and its 
stochastic representation Eq.~(\ref{eqNonl}) is in good 
agreement with the theoretical quadratic nonlinearity prediction
and the experimental measurements. Indeed it has been pointed out that it is sufficient to fit the plasma density signals 
using a limiting beta distribution which coincides with the gamma PDF for positive
skewness and using a shifted gamma r.v., $2\left<\psi\right>-\psi$, for negative skewness \cite{Labit}. 
The comparison between the gamma and the beta processes shows the importance of the PDF functional form used
to fit experimental data. With identical two first moments, different distributions 
(gamma, beta and log-normal) could provide 
a reasonable fit of experimental measurements, although the nonlinearity degrees of their  dynamics is different.  
Therefore,
 and in order to improve the connection between the theory of nonlinear dynamics and the experimental time series,
the analysis of plasma's fluctuations using  strong criteria \cite{Kagan,Lukacs-55,McKenzie} are suitable to clarify
how long the statistics follows a given distribution. 
This procedure is preferable to the data fitting and a statistical signature based on the $K$-$S$ scaling, 
since no unicity exists between this scaling and the underlying distributions.  \\
\indent As a summary, we have derived a quadratically nonlinear stochastic differential equation for 
the intermittent plasma density in the SOL of fusion devices. The plasma density dynamics evolves 
on a turbulence correlation time scale ($\tau_c=1-100$ $\mu$s),
 and is characterized by the local fluctuations amplitude $R=10-90\%$. 
This equation is consistent with the experimental measurements and a theoretical prediction,
 since it behaves a gamma marginal and a quadratic nonlinearity.
Its simple representation is suitable for edge plasma modeling, since a finite  plasma density correlation time, 
together with high fluctuations amplitude affect
considerably the plasma-wall interaction and the associated transport
of sputtered impurities and released molecules \cite{Pigarov_12,Samad_12}.
The sensitivity of the nonlinearity degrees to the intermittent distribution has been 
investigated by comparing two potentially used distributions. Our result shows that the gamma distribution agrees with 
the quadratic nonlinearity, in comparison with the beta distribution leading to a root square nonlinearity.\\



\end{document}